\documentclass[epj]{svjour}
\usepackage{bbding}
\usepackage{amssymb}
\usepackage{amsmath}
\usepackage{graphics}
\begin{document}
\title{State transfer in intrinsic decoherence spin channels}

\author{Ming-Liang Hu\inst{1}\thanks
{E-mail: mingliang0301@xupt.edu.cn} and Han-Li Lian\inst{2}
}
\institute{Department of Applied Mathematics and Applied Physics,
Xi'an Institute of Posts and Telecommunications, Xi'an 710061, P.R.
China }
\date{Received: date / Revised version: date}
%
\abstract{ By analytically solving the master equation, we
investigate quantum state transfer, creation and distribution of
entanglement in the model of Milburn's intrinsic decoherence. Our
results reveal that the ideal spin channels will be destroyed by the
intrinsic decoherence environment, and the detrimental effects
become severe as the decoherence rate $\gamma$ and the spin chain
length $N$ increase. For infinite evolution time, both the state
transfer fidelity and the concurrence of the created and distributed
entanglement approach steady state values, which are independent of
the decoherence rate $\gamma$ and decrease as the spin chain length
$N$ increases. Finally, we present two modified spin chains which
may serve as near perfect spin channels for long distance state
transfer even in the presence of intrinsic decoherence environments
$\mathcal {F}{[\rho(t)]}$.
} 
\maketitle
\section{Introduction}
\label{intro} In quantum information processing (QIP), it is
desirable to achieve a high-fidelity transfer of quantum states
between different parts, such as the core processor, storage, etc.,
of a quantum computer. To this end, a variety of solid-state spin
networks with always-on interactions have been proposed [1-16].
Particularly, Christandl et al. showed that with elaborately
designed modulated exchange couplings between neighboring spins, one
can implement perfect quantum state transfer (QST) over arbitrary
distances between the opposite ends of a XX spin chain or between
the two antipodes of the one-link and the two-link hypercubes with
however the maximum perfect communication distance $\rm2log_{3}\it
N$ [3,4]. In addition, they also showed that these modulated spin
structures can distribute arbitrary entanglement between two distant
parties. Zhang and Long et al. [13] realized this perfect state
transfer algorithm in a three-qubit XX chain using liquid NMR
system. Later, Shi et al. presented a class of more general
pre-engineered perfect spin channels [6] according to the
spectrum-parity-matching condition (SPMC) they deduced. Then, Kostak
et al. [14] established a general formalism for engineering spin
Hamiltonians for perfect state transfer in networks of arbitrary
topology and coupling configuration. Christandl's innovative works
were extended by Jafarizadeh and Sufiani in a recent work [15], in
which they adopted distance-regular graphs as spin networks and
found that any such network (not just the hypercube) can achieve
unit fidelity of state transfer over arbitrarily long distances.
Moreover, D'Amico et al. [10] showed that one can create and
distribute entanglement with an interaction-modulated Y-shaped spin
network, particularly, with a slightly complicated bifurcation
structure, the distributed entanglement can be frozen when a phase
flip is applied to one spin out of each pair.

In addition to the above-mentioned protocols which mainly
concentrated on spin chains with nearest-neighbor (NN) couplings, in
Ref. [17] Paternostro et al. studied QST in imperfect artificial
spin networks with all the qubits are mutually coupled (in which the
usually assumed NN coupling is invalid). They presented a strategy
to avoid the spoiling effects of these redundant connections with a
modification of the couplings of the first and the last qubits in
the chain, which enables nearly optimal state transfer. Then in Ref.
[18] Kay demonstrated that perfect state transfer is also possible
in the presence of next-nearest-neighbor (NNN) couplings. Moreover,
compared to the case where the system contains only two-spin
interactions, the authors in Ref. [19] presented a scheme of QST by
introducing the three-spin interaction, and showed that they can
significantly increase the speed of QST in an XY chain. Besides the
spin-half systems, state and entanglement transfer driven by a
bilinear-biquadratic (BB) spin-1 Heisenberg chain was also discussed
recently [20], in which the authors concentrated on the relations
between the transfer efficiency and the quantum phase transitions.

Most recently, a milestone work appears in Ref. [21] presented a
control-limited scheme [22] for perfect state transfer through a
pre-engineered spin chain with the help of local end-chain
single-qubit operations. While nearly all of the previous schemes
whose achievements of perfect state transfer relies crucially on the
preparation of the spin medium in a fiducial pure state, the authors
in Ref. [21] demonstrated that state initialization of the spin
medium is inessential to the performance of the protocol if proper
encoding at the end of the chain is performed. The key requirements
for their scheme are the arrangement of proper time evolution and
the performance of clean projective measurements on the two end
spins. This innovative work considerably relaxes the prerequisites
for obtaining reliable QST across interacting-spin systems.
Stimulated by this innovative work, in Ref. [23] Markiewicz and
Wie\'{s}niak proposed a special type of two-qubit encoding strategy
for perfect state transfer, where no remote-cooperated global state
initialization and any additional communication are needed.

Apart from these exciting progresses, we noted that although there
are several works [24-36] concerning the decoherence effects on
entanglement dynamics, studies thus far has seldom consider the
influence of different kinds of decoherence scenarios on transfer of
quantum states due to the complex and unclear mechanism of its
interaction with the environments. However, from a practical point
of view, all the real physical systems, especially a solid-state
system, will unavoidably be influenced by its surrounding
environments. This influence can cause the initial state of the
system of interest becomes entangled with the environment in an
uncontrollable way, and it is just this entanglement of the system
with the environment that causes decoherence. The decoherence can
greatly affects the transfer efficiency of quantum states, as well
as generation and distribution of entanglement, and thus becomes one
of the dominating obstacles baffling the physical implementation of
QIP. It is therefore of great importance and fundamentally
interesting to find ways to prevent or minimize the detrimental
effects in the practical realization of QIP.

The standard way to investigate decoherence is to consider the
system of interest as a part of a larger closed system involving the
environment, and the density operator of the system can then be
obtained by tracing out all other degrees except quantum states of
the system. In the present paper, however, we would like to resort
to a different approach, i.e., the scenario of the so-called
intrinsic decoherence proposed by Milburn [37], who modified the
Schr\"{o}dinger equation in such a way that quantum coherence is
automatically destroyed as the system evolves. Such a consideration
is fed by two motivations. First, this model is amenable to exact
analytical treatment as we will see, one can determine the density
operator of the system at arbitrary time $t$ by the sole knowledge
of the eigenvalues and eigenvectors of the system. Second, although
the absence of unitarity for a closed system in this model makes it
unlikely to be a fundamental description of decoherence, its
stochastic behavior in time evolution may still be an effective
approximation for describing the phenomenon of the system. For
example, it has been applied to describe decoherence of a single
trapped ion due to intensity and phase fluctuations in the exciting
laser pulses [38]. Dynamics of the mutual entropy of two-coupled
Josephson charge qubits with intrinsic decoherence has also been
studied recently [39]. Moreover, as pointed by the authors of Ref.
[29,40], this model may be available in approximately describing the
non-dissipative decoherence of several physical systems in the
presence of white noise.

\section{General formalism}
\label{sec:2} In this paper, we consider quantum state transfer
properties in the model of Milburn's intrinsic decoherence [37]. The
kernel of this decoherence scenario is the postulate that on
sufficiently short time steps the system does not evolve
continuously under unitary evolution but rather in a stochastic
sequence of identical unitary transformation, which can account for
the disappearance of quantum coherence as the system evolves. Based
on this assumption, Milburn obtained the master equation (in units
of $\hbar$) governing the time evolution of the system
\begin{equation}
{d\rho\over dt}={1\over\gamma} {[\exp(-i\gamma \hat{H})\rho
\exp(i\gamma \hat{H})-\rho]},
\end{equation}
where $\gamma$ is the intrinsic decoherence parameter (the mean
unitary time step). Expanding Eq. (1) to the first order in
$\gamma$, one finds
\begin{equation}
{d\rho\over
dt}={-i[\hat{H},\rho]}-{\gamma\over2}[\widehat{H},[\hat{H},\rho]].
\end{equation}

The first term on the right-hand side of Eq. (2) generates a
coherent unitary time evolution of the system, while the second
term, which does not commute with the Hamiltonian, represents the
decoherence effect on the system and generates an incoherent
dynamics of the system. In the limit of $\gamma\rightarrow 0$, the
ordinary Schr\"{o}dinger equation is recovered.

To solve Eq. (2), one can define three auxiliary superoperators
$\hat J$, $\hat S$ and $\hat L$, which satisfy
\begin{equation}
\hat J\rho=\gamma\hat H\rho\hat H,\quad\hat S\rho=-i[\hat H,
\rho],\quad \hat L\rho=-\frac{\gamma}{2}\{\hat H^{2},\rho\}.
\end{equation}

From Eq. (3) it is straightforward to show that
\begin{eqnarray}
\exp(\hat J\tau)\rho(t)&=&\sum_{l=0}^{\infty}\frac{(\gamma
                          \tau)^{l}}{l!}\hat H^l\rho(t)\hat H^l, \nonumber\\
\exp(\hat S\tau)\rho(t)&=&\exp(-i\hat H\tau)\rho(t)\exp(i\hat H\tau), \nonumber\\
\exp(\hat L\tau)\rho(t)&=&\exp\left(-\frac{\gamma\tau}{2}\hat
                         H^2\right)\rho(t)\exp\left(-\frac{\gamma\tau}{2}\hat H^2\right).
\end{eqnarray}

Thus Eq. (2) simplifies to $d\rho/dt=(\hat J+\hat S+\hat L)\rho$,
and its formal solution can be written in terms of the Kraus
operators $\hat M _{l}(t)$ as
\begin{equation}
\rho(t)=\sum_{l=0}^{\infty}\hat M_{l}(t)\rho(0)\hat M_{l}^{\dag
}(t),
\end{equation}
where $\rho(0)$ denotes the initial state of the system, $\hat
M_{l}(t)=(\gamma t)^{l/2}\hat H^l\exp(-i\hat Ht)\exp(-\gamma t\hat
H^2/2)/\sqrt{l!}$ satisfies the relation
 $\sum_{l=0}^{\infty}\hat{M}_l^{\dag}(t)\hat{M}_l(t)=1$ for all time
$t$.

If we rewrite $\rho(0)$ in forms of the energy eigenstate basis as
$\rho(0)=\sum_{kk'}a_{kk'}|\psi_{k}\rangle\langle \psi_{k'}|$, then
we obtain
\begin{equation}
\rho(t)=\sum_{kk'}a_{kk'}\exp[-it(E_k-E_{k'})-\frac{\gamma
t}{2}(E_k-E_{k'})^2]|\psi_{k}\rangle\langle \psi_{k'}|,
\end{equation}
where $a_{kk'}=\langle\psi_k|\rho(0)|\psi_{k'}\rangle$, $E_k$ and
$|\psi_k\rangle$ are eigenvalue and the corresponding eigenvector of
the considered system.

For the special case that $\rho(0)$ is an eigenstate of the system,
$a_{kk'}\neq0$ only when $k=k'$, Thus from Eq. (6) one can obtain
$\rho(t)=\sum_{k}a_{kk}|\psi_k\rangle\langle\psi_k|\equiv\rho(0)$,
the system will be unaffected by the intrinsic decoherence during
the time evolution process.

Furthermore, for a spin chain Hamiltonian commutes with the total
$z$ component of the spin, i.e., $[\hat H,\sigma_{\rm tot}^{z}]=0$,
where $\sigma_{\rm tot}^{z}=\sum_{i}\sigma_{i}^{z}$, the
$2^N\otimes2^N$ Hilbert space can be decomposed into $N+1$ different
invariant subspaces, each of which is a distinct eigenspace of the
operator $\sigma_{\rm tot}^{z}$, and a system prepared in these
subspaces will remains in them. In the single-excitation invariant
subspace $\mathcal{H}_{\rm 1}$ spanned by the site basis
$|n\rangle=\sigma_n^+|0\rangle^{\otimes N}$ $(n=1, 2, ..., N)$, one
can rewrite $\rho(t)$ as $\rho(t)=\sum_{nm}b_{nm}|n\rangle\langle
m|$, then in the standard basis $\{|00\rangle, |01\rangle,
|10\rangle, |11\rangle\}$, the single qubit reduced density matrix
can be obtained as
\begin{equation}
\rho_{i}(t)=\left(\begin{array}{cc}
 1-b_{ii}& 0 \\
 0       & b_{ii}
\end{array}\right).
\end{equation}

Similarly, one can obtain the two-qubit reduced density matrix
between qubits $i$ and $j$ as
\begin{equation}
\rho_{ij}(t)= \left(\begin{array}{cccc}
 1-b_{ii}-b_{jj}& 0& 0& 0 \\
 0 & b_{jj}&b_{ji}&0 \\
 0&b_{ij}&b_{ii}&0 \\
 0&0&0&0
\end{array}\right).
\end{equation}

In this paper, we use the fidelity
$F=\langle\psi(0)|\rho_i(t)|\psi(0)\rangle$ as an estimation of the
quality of the state transfer from the sender to the destination
qubits [1], and adopt the concept of concurrence $C=\rm{max}\{0,
\lambda_1-\lambda_2-\lambda_3-\lambda_4\}$ as a measure of the
pairwise entanglement [31,32]. Here the quantities $\lambda_i$
$(i=1, 2, 3, 4)$  are the square roots of the eigenvalues of the
product matrix
$R=\rho(\sigma^y\otimes\sigma^y)\rho^*(\sigma^y\otimes\sigma^y)$ in
decreasing order.

From Eqs. (7), (8) and the above definitions about transfer fidelity
and concurrence, one can obtain directly that $F(N,t)=|b_{NN}|$ and
$C_{ij}(N,t)=2|b_{ij}|$ for a state initially prepared in the
$N$-dimensional subspace $\mathcal {H}_{\rm1}$.

Another quantity related to the efficiency of the quantum spin
channel of interest is the fidelity averaged over all pure states in
the Bloch sphere. The state of the whole system at the initial time
$t=0$ can be written as

\begin{equation}
|\psi(0)\rangle=\cos{\frac{\theta}{2}}|\textbf{0}\rangle+e^{i\phi}\sin{\frac{\theta}{2}}|s\rangle,
\end{equation}
where $|\textbf{0}\rangle=|00...0\rangle$,
$|s\rangle=\sigma_s^+|0\rangle^{\otimes N}$, $\theta$ and $\phi$ are
arbitrary phase angles.

For this type of initial state, its dynamics is completely
determined by the evolution in the zero and single excitation
subspace $\mathcal {H}_{\rm0\oplus1}$. From Eq. (6) one can obtain
the state at time $t$ as
\begin{eqnarray}
\rho(t)&=&\cos^2\frac{\theta}{2}|\textbf{0}\rangle\langle\textbf{0}|+\sin^2\frac{\theta}{2}\sum_{n,m=1}^{N}a_{nm}|n\rangle\langle
m|\nonumber\\&& +\left(e^{i\phi}\sin\frac{\theta}{2}
\cos\frac{\theta}{2}\sum_{n=1}^{N}b_{n}|n\rangle\langle\textbf{0}|+\rm
H.c.\right),
\end{eqnarray}
with the coefficients $a_{nm}$ and $b_n$ given by
\begin{eqnarray}
a_{nm}&=&\sum_{k,k'=1}^{N}c_{k,s}c_{k',s}c_{k,n}c_{k',m}\times\nonumber\\&&
         \exp\left[-it(E_k-E_{k'})-\frac{\gamma t}{2}(E_k-E_{k'})^2\right],\nonumber\\
b_{n}&=&\sum_{k=1}^{N}c_{k,s}c_{k,n}\exp\left(-itE_k-\frac{\gamma
t}{2}E_k^2\right),
\end{eqnarray}
where $c_{k,n}$ is the amplitude of coefficient for the state
$|n\rangle$ in the eigenstate
$|\tilde{k}\rangle=\sum_{n=1}^{N}c_{k,n}|n\rangle$.

Then by tracing off the states of all other spins except $i$ from
$\rho(t)$, one has
\begin{equation}
\rho_{i}(t)=\left(\begin{array}{cc}
 1-a_{ii}\sin^2\frac{\theta}{2}& b_i^*e^{-i\phi}\sin\frac{\theta}{2}\cos\frac{\theta}{2} \\
 b_ie^{i\phi}\sin\frac{\theta}{2}\cos\frac{\theta}{2} & a_{ii}\sin^2\frac{\theta}{2}
\end{array}\right).
\end{equation}

From Eqs. (9), (12), the fidelity
$F=\langle\psi(0)|\rho_i(t)|\psi(0)\rangle$ can be obtained as
\begin{eqnarray}
F&=&\cos^2\frac{\theta}{2}\left(1-a_{ii}\sin^2\frac{\theta}{2}+2|b_i|\sin^2\frac{\theta}{2}\cos\alpha\right)+\nonumber\\&&
a_{ii}\sin^4\frac{\theta}{2},
\end{eqnarray}
where $\alpha=\rm{arg}\it(b_i)$ denotes the argument of the complex
number $b_i$.

Thus the average fidelity $\bar
F=\frac{1}{4\pi}\int\langle\psi(0)|\rho_i(t)|\psi(0)\rangle d\Omega$
can be calculated as
\begin{equation}
\bar F=\frac{|b_i|\cos(\alpha)}{3}+\frac{a_{ii}}{6}+\frac{1}{2}.
\end{equation}

From Eq. (11) one can see that in the absence of intrinsic
decoherence (i.e., $\gamma=0$), the equality $a_{ii}=|b_i|^2$ holds,
thus Eq. (14) reduces to Eq. (6) in Ref. [1], which describes
average fidelity in the non-disturbed case.

\section{State transfer in decoherence spin channels}
\label{sec:3} We first consider quantum state transfer via spin
chain governed by the XX Hamiltonian
\begin{eqnarray}
\hat{H}=\frac{J}{2}\sum_{n=1}^{N-1}(\sigma_n^x\sigma_{n+1}^x+\sigma_n^y\sigma_{n+1}^y),
\end{eqnarray}
where $\sigma_n^\alpha$ $(\alpha=x, y, z)$ are the usual Pauli
matrices of the $n$th qubit.

For this model, Christandl et al. have shown that perfect state
transfer from one end of the chain to another is only possible for
the case of chain length $N=2$ and $N=3$, respectively [3,4]. Here
we show that this ideal communication channel will be destroyed
under the influence of intrinsic decoherence.

The eigenvalues and eigenvectors of the Hamiltonian (15) can be
obtained as
\begin{eqnarray}
E_{k}&=&2J\cos\frac{\pi k}{N+1},\nonumber\\
|\tilde{k}\rangle&=&\sqrt{\frac{2}{N+1}}\sum_{n=1}^{N}\sin\left(\frac{\pi
kn}{N+1}\right)|n\rangle.
\end{eqnarray}

We first consider transfer of an excitation across the chain. For
this purpose, we assume the system is initially prepared in the
state $|n_0\rangle$. In the energy eigenstate basis, $|n_0\rangle$
can be expressed as
\begin{eqnarray}
|n_0\rangle=\sqrt{\frac{2}{N+1}}\sum_{k=1}^{N}\sin\left(\frac{\pi
kn_0}{N+1}\right)|\tilde{k}\rangle.
\end{eqnarray}
Thus one has
\begin{eqnarray}
\rho(0)=\frac{2}{N+1}\sum_{k,k'=1}^{N}\sin\left(\frac{\pi
kn_0}{N+1}\right)\sin\left(\frac{\pi
k'n_0}{N+1}\right)|\tilde{k}\rangle\langle \tilde{k}'|.\nonumber\\
\end{eqnarray}
Combination of Eqs. (6) and (18) gives rise to
\begin{eqnarray}
\rho(t)&=&\frac{4}{(N+1)^2}\sum_{n,m=1}^{N}\sum_{k,k'=1}^{N}\sin\left(\frac{\pi
kn}{N+1}\right)\sin\left(\frac{\pi k'm}{N+1}\right) \nonumber\\&&
\times\sin\left(\frac{\pi kn_0}{N+1}\right)\sin\left(\frac{\pi
k'n_0}{N+1}\right)\times \nonumber\\&&
\exp\left[-i2Jt\left(\cos{\frac{\pi k}{N+1}}- \cos{\frac{\pi
k'}{N+1}}\right)\right]\times\nonumber\\&& \exp\left[-2J^2\gamma
t\left(\cos{\frac{\pi k}{N+1}}- \cos{\frac{\pi
k'}{N+1}}\right)^2\right]|n\rangle \langle m|.\nonumber\\
\end{eqnarray}

For initial state $|1\rangle$ prepared in the input node A, the
transfer fidelity of the output state in node B can be obtained from
Eq. (19), and typical plots for the cases of $N=2$ and $N=3$ with
different decoherence rates are shown in Fig. 1, where the coupling
constant $J$ is chosen to be 1. In big contrast to the ideal case
(i.e., $\gamma=0$), one can see that the transfer fidelity $F$
behaves as a damped oscillation as the time $t$ evolves. This
phenomenon can be understood from Eq. (19), where the product of the
first five terms on the right-hand side causes the oscillations, and
the last term introduces the amplitude damping. With the increase of
the decoherence rate $\gamma$, or the chain length $N$, the
detrimental effects becomes more severe and therefore more quantum
state information will be lost. Thus for spin networks with
identical neighboring qubit couplings, even if for the one-link and
two-link hypercube geometries, perfect transfer of an excitation is
still impossible in the intrinsic decoherence environments.
\begin{figure}
\centering
\resizebox{0.4\textwidth}{!}{%
\includegraphics{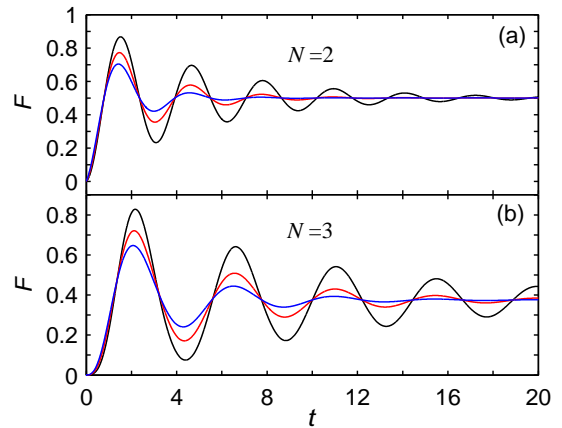}}
\caption{(Color online) Dynamics of the state transfer fidelity $F$
for the XX chain with identical interactions. The decoherence rate
is given by $\gamma=0.1$ (black), $\gamma=0.2$ (red) and
$\gamma=0.3$ (blue).} \label{fig:1}
\end{figure}

For infinite time $t$, the system evolves into a steady state with
the transfer fidelity arrives at an asymptotic value $F^{\rm
steady}(N)$, which can be obtained by combination of Eqs. (7), (19)
and taking the infinite-time limit. After a tedious computation, we
obtain
\begin{eqnarray}
F^{\rm steady}(N)=\frac{3}{2(N+1)}.
\end{eqnarray}

Clearly, this steady state transfer fidelity is independent of the
decoherence rate $\gamma$, and it solely decreases with the increase
of the chain length $N$.

Next we consider time-dependence of the average fidelity for the XX
spin chain with identical interactions and subject to intrinsic
decoherence environments, with initial state prepared in the form of
Eq. (9) in node A, i.e., $s=1$. From Fig. 2 one can see clearly that
the average fidelity $\bar F$ also behaves as a damped oscillation
as the time evolves. Here the relative small value for the case of
$N=2$ is due to the fact that the phase of the state at node B is
uncorrected, i.e., $\alpha$ is not a multiple of $2\pi$. When
$t\rightarrow \infty$, the average fidelity also arrives at a steady
state value, which is independent of the decoherence rate $\gamma$,
and can be obtained analytically by taking the infinite-time limit
of $\bar F$ from Eqs. (11), (14), and (16) as
\begin{equation}\label{eq:21}
\bar{F}^{\rm steady}(N)=\left\{
    \begin{aligned}
         &\frac{6N+17}{12N+12}\quad\rm if\quad\it N\in\rm odd,\\
         &\frac{2N+3}{4N+4}\quad\quad\rm if\quad\it N\in\rm even.
    \end{aligned} \right.
\end{equation}

Contrary to that of the initial state $|1\rangle$, this steady value
does not decrease monotonously with the increase of the chain length
$N$. However, as can be seen from Eq. (21), they decrease with the
increase of the odd and even $N$, respectively, and approach to the
asymptotic value 0.5 in the limit of $N\rightarrow \infty$.
\begin{figure}
\centering
\resizebox{0.4\textwidth}{!}{%
\includegraphics{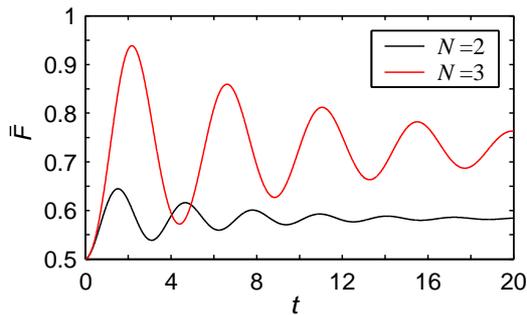}
} \caption{(Color online) Dynamics of the average fidelity $\bar{F}$
for the XX spin chain with identical interactions, where the
decoherence rate is given by $\gamma=0.1$.} \label{fig:2}
\end{figure}

In the following we discuss quantum state transfer in intrinsic
decoherence spin channels with fixed but different couplings between
qubits. We consider the following modified Hamiltonian
\begin{eqnarray}
\hat{H}=\sum_{n=1}^{N-1}\frac{J_{n,n+1}}{2}(\sigma_n^x\sigma_{n+1}^x+\sigma_n^y\sigma_{n+1}^y),
\end{eqnarray}
where $J_{n,n+1}=\lambda\sqrt{n(N-n)}$ is the modulated exchange
coupling, and $\lambda$ is a scaling constant.

The above Hamiltonian is identical to the representation of the
Hamiltonian $\hat{H}_s$ of a fictitious spin $S=(N-1)/2$ particle:
$\hat{H}_s=\lambda S_x$, where $S_x$ is its angular momentum
operator in $x$-direction and $\lambda$ is a scaling constant. For
this Hamiltonian, its eigenvalues and corresponding eigenvectors can
be obtained as [43]
\begin{eqnarray}
E_{k}=(-N+2k-1)\lambda,\quad
|\tilde{k}\rangle=\sum_{n=1}^Nc_{k,n}|n\rangle.
\end{eqnarray}
where the coefficient $c_{k,n}$ is given by the following recursion
relations
\begin{eqnarray}
c_{1,1}&=&1/2^{(N-1)/2}, c_{k,1}=(-1)^{k+1}c_{1,k}\nonumber\\
c_{k,n}&=&\frac{2E_kc_{k,n-1}-\sqrt{(n-2)(N-n+2)}c_{k,n-2}}{\sqrt{(n-1)(N-n+1)}}
\quad(n\geqslant 2).\nonumber\\
\end{eqnarray}

\begin{figure}
\centering
\resizebox{0.4\textwidth}{!}{%
\includegraphics{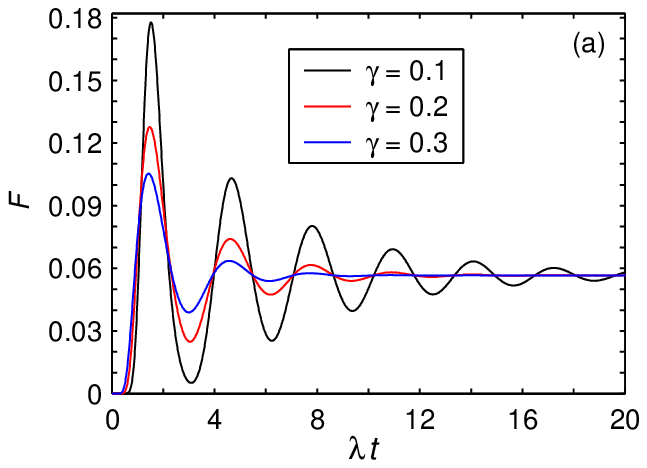}}
\resizebox{0.4\textwidth}{!}{%
\includegraphics{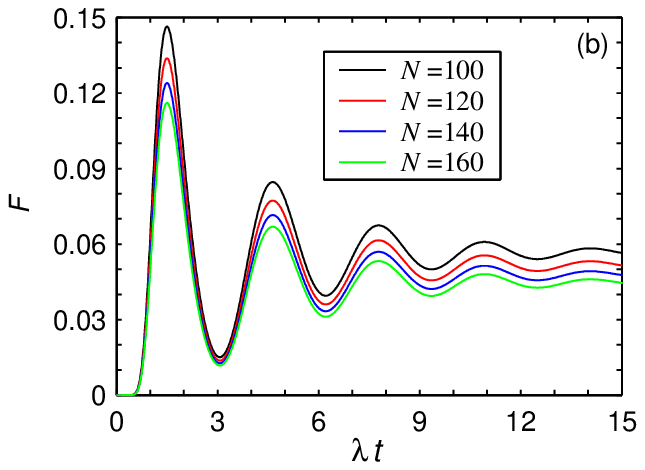}
} \caption{(Color online) Dynamics of the state transfer fidelity
$F$ for the XX chain with modulated interactions. (a) chain length
$N=100$ with different decoherence rate; (b) decoherence rate
$\gamma=0.15$ with different chain length.} \label{fig:3}
\end{figure}

For this modulated chain, it has been shown that one can achieve
perfect state transfer between the input node $n$ and the output
node $N-n+1$ after a time $t_0=\pi/2\lambda$ and at intervals of
$\pi/\lambda$ thereafter in the absence of decoherence environment
[3,4]. When the intrinsic decoherence is present, however, this
ideal spin channel will be destroyed, and it acts as an amplitude
damping quantum channel as the rescaled time $\lambda t$ evolves. As
can be seen from Fig. 3, the transfer fidelity $F$ oscillates around
a steady state value, with the amplitude decreases gradually. This
detrimental effects becomes more and more severe with the increase
of the decoherence rate and the spin chain length, which is in
consistent with the cases of the two- and three-site spin chains
with identical interactions (In fact, they are two special cases of
the interaction-modulated spin chain). This puts new constraints on
these spin chains for long distance quantum state transfer. When
$t\rightarrow \infty$, the transfer fidelity reaches a steady state
value, which can be obtained from Eqs. (6), (7), (23), and (24) as
\begin{eqnarray}
F^{\rm{steady}}(N)=\frac{1}{2^{2N-2}}\prod_{k=2}^N\left(4-\frac{2}{k-1}\right).
\end{eqnarray}

The steady state transfer fidelity of the interaction-modulated spin
chain is still independent of the decoherence rate $\gamma$, and its
magnitude is larger than its unmodulated counterparts [cf. Eqs. (20)
and (25)], thought it still decreases with the increase of the chain
length $N$.
\begin{figure}
\centering
\resizebox{0.4\textwidth}{!}{%
\includegraphics{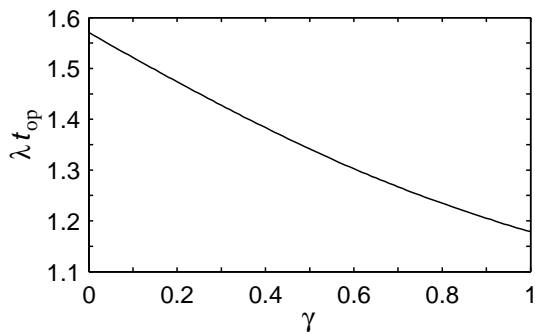}
} \caption{Dependence of $\lambda t_{\rm op}$ on decoherence rate
$\gamma$. Note that the magnitudes of $\lambda t_{\rm op}$ is
independent of the chain length $N$.} \label{fig:4}
\end{figure}

On the other hand, since the detrimental effects become severe as
the rescaled time $\lambda t$ evolves, one may expect there exists
an optimal time $\lambda t_{\rm op}$ at which the state transfer
fidelity $F$ gets its maximum value. In Fig. 4 we show $\lambda
t_{\rm op}$ versus the intrinsic decoherence rate $\gamma$, from
which one can see that $\lambda t_{\rm op}$ is shifted to the
left-hand side of $\lambda t_0=\pi/2\simeq 1.57$, and it decreases
with the increase of $\gamma$. Our numerical results also revealed
that the magnitudes of $\lambda t_{\rm op}$ is independent of the
chain length $N$.
\begin{figure}
\centering
\resizebox{0.4\textwidth}{!}{%
\includegraphics{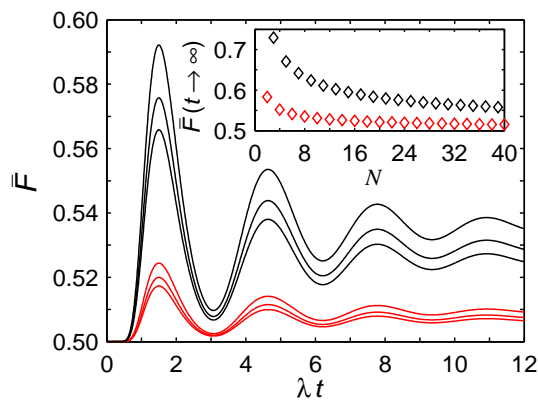}
} \caption{(Color online) Dynamics of the average fidelity $\bar{F}$
for the XX chain with modulated interactions, where the decoherence
rate $\gamma=0.15$. The black lines from top to bottom correspond to
chain length $N=101$, 151, and 201; whereas the red lines from top
to bottom correspond to chain length $N=100$, 150, and 200. The
inset shows the asymptotic value of the average fidelity for
infinite time $t$ versus chain length $N$.} \label{fig:5}
\end{figure}

When considering the average fidelity, the numerical results
calculated from Eqs. (11), (14), (23), and (24) show that it
displays qualitatively the similar behaviors with that displayed in
Fig. 3. The average fidelity decreases with increasing value of both
odd and even $N$, respectively, and the chain with odd-number qubits
seems to be more robust on creating high-fidelity state transfer in
the presence of intrinsic decoherence (see Fig. 5). Moreover, as can
be seen from the inset of Fig. 5, the average fidelity goes to a
steady state value in the limit of $t\rightarrow \infty$, which has
no relation with the decoherence rate $\gamma$. They decrease with
the increase of both odd and even $N$, and approach the asymptotic
value 0.5 in the limit of $N\rightarrow \infty$.

In the absence of intrinsic decoherence (i.e., $\gamma=0$), the
above interaction-modulated spin chain can also be used to perfectly
transfer an entangled state from one end of the chain to another
[4]. When the decoherence is present, however, this ideal spin
channel will be destroyed. For example, If one start with the Bell
state $|\psi^\pm\rangle=(|01\rangle\pm|10\rangle)/\sqrt{2}$ on the
first two qubits of the chain, the temporal evolution of the
concurrence will behaves similarly as the state transfer fidelity,
i.e., it acts as an amplitude damping channel. When the rescaled
evolution time $\lambda t$ approaches infinite, from the formulae
described in Section 2 one can obtain
\begin{eqnarray}
C_{1,2}^{\rm {steady}}(N)=C_{N-1,N}^{\rm
{steady}}(N)=\prod_{n=3}^{N}\frac{2n-5}{2n-4}.
\end{eqnarray}

In fact, one can show that for the initial state
$|\psi\rangle=a|01\rangle\pm b|10\rangle$ $(|a|^2+|b|^2=1)$ prepared
on the first two qubits, the following relation holds
\begin{eqnarray}
C_{1,2}^{\rm {steady}}(N)=C_{N-1,N}^{\rm {steady}}(N)=C_{1,2}^{\rm
{initial}}\prod_{n=3}^{N}\frac{2n-5}{2n-4},
\end{eqnarray}
where $C_{1,2}^{\rm {initial}}=2|ab|$ denotes the concurrence of the
initial state of the first two qubits. This indicates that when the
rescaled evolution time $\lambda t$ approaches infinite, the system
goes to a steady mirror-symmetric state with two-qubit reduced
density matrix $\rho_{nm}(t)=\rho_{N-n+1,N-m+1}(t)$, and the steady
state value $C_{N-1,N}^{\rm {steady}}(N)=C_{1,2}^{\rm {steady}}(N)$
decreases as the chain length $N$ increases.

We now investigate entanglement distribution between two distant
parties through the intrinsic decoherence spin channel. For this
purpose, we assume the entangled state
$|\psi\rangle=(|01\rangle+|10\rangle)/\sqrt{2}$ is initially
prepared between a noninteracting qubit NI and the first qubit A on
the chain, then after some time $t$, the entanglement will be
established between NI and the target spin B. The overall
Hamiltonian of the system can be written as
$\hat{H}'=I\otimes\hat{H}$, and with the same method used above, one
can demonstrate that the concurrence $C_{\rm{NI,B}}(N,t)$ (Note that
here $N$ denotes the length of the interacting-spin chain, and does
not include the noninteracting qubit NI) also behaves as a damped
oscillation, and when $t\rightarrow\infty$, we obtain
\begin{eqnarray} \label{eq:28}
C_{\rm{NI,A}}^{\rm{steady}}(N)&=&C_{\rm{NI,B}}^{\rm{steady}}(N)\nonumber\\
&=&\left\{\begin{aligned}
 &0\quad\rm \quad\quad\quad\quad\quad\quad if\quad\it N\in\rm odd, \\
 &\prod_{n=3}^{(N+4)/2}\frac{2n-5}{2n-4}\quad\rm if\quad\it N\in \rm
 even.
\end{aligned} \right.
\end{eqnarray}

This equation shows clearly that the XX chain with even-number
qubits is more robust than its counterpart with odd-number qubits on
distributing quantum entanglement. This is somewhat different from
that of the average fidelity (see Fig. 5), where the chain with
odd-number qubits is more efficient on creating high-fidelity state
transfer in the presence of intrinsic decoherence.

\section{Creating entanglement in decoherence environments}
\label{sec:4} In this section, we see intrinsic decoherence effects
on the creation of entanglement in various kinds of spin networks.
For this purpose, we consider the multiarm structure
$M(l_1,l_2,N_A)$ of the XX Hamiltonian (22) with the addition of the
exchange couplings between the hub site and its nearest-neighbor
output sites satisfy the branching rule [10]. Here $l_1$ and $l_2$
denote the number of sites in the input and output arms,
respectively, and $N_A$ is the number of output arms (see Fig. 6).
It has been shown that in the absence of decoherence environment,
this structure can be employed to create multi-qubit entangled $W$
state at the ends of the outgoing arms.
\begin{figure}
\centering
\resizebox{0.35\textwidth}{!}{%
\includegraphics{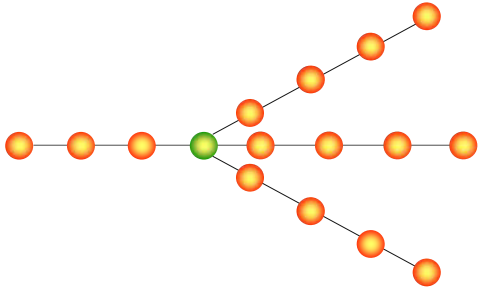}}
\caption{(Color online) Sketch of the multiarm structure of the spin
network, where the green circle denotes the hub. The number of sites
in the input and output arms are $l_1=3$, $l_2=4$, and the number of
output arms is $N_A=3$.}
\label{fig:6}       
\end{figure}

The Hamiltonian in the subspace $\mathcal {H}_1$ spanned by the
basis vectors $|n\rangle$ ($n=1, 2,... , N$ ) is
\begin{eqnarray}
\hat{H}=\sum_{(ij)}J_{ij}|i\rangle\langle j|+\rm H.c.,
\end{eqnarray}
where the summation runs over all pairs of neighboring spins. For
the sake of simplicity, we first consider the Y-shaped structure
$Y(l_1,l_2,2)$. The total number of sites now is $N=l_1+2l_2+1$. To
examine temporal evolution of the concurrence of the prepared
initial state, we make the following basis transformation for spins
just in the same position of each arm
\begin{eqnarray}
|n^\pm\rangle=\frac{1}{\sqrt {2}}(|n\rangle\pm|n'\rangle),
\end{eqnarray}
where $n, n'>l_1+1$. Then in the subspace spanned by $|n\rangle$
$(n\leqslant l_1+1)$ and $|n^\pm\rangle$ $(n>l_1+1)$, the
Hamiltonian can be rewritten as
\begin{eqnarray}
\hat{H}&=&\sum_{n<l_1+1}J_{n,n+1}|n\rangle\langle n+1|+\nonumber\\&&
\sum_{n>l_1+1}\sum_{r=+,-}J_{n,n+1}|n^r\rangle\langle
(n+1)^r|+\nonumber\\&& \sqrt{2}J_{l_1+1,l_1+2}|l_1+1\rangle\langle
(l_1+2)^+|+\rm H.c.
\end{eqnarray}

Clearly, under the transformation (30) the Y-shaped structure is
transformed into a linear chain consisting of the input arm, the hub
and one output arm while the other output arm is decoupled (see Fig.
7), i.e., this structure is identical to the interaction-modulated
one-dimensional XX chain with chain length $l=l_1+l_2+1$.
\begin{figure}
\centering
\resizebox{0.35\textwidth}{!}{%
\includegraphics{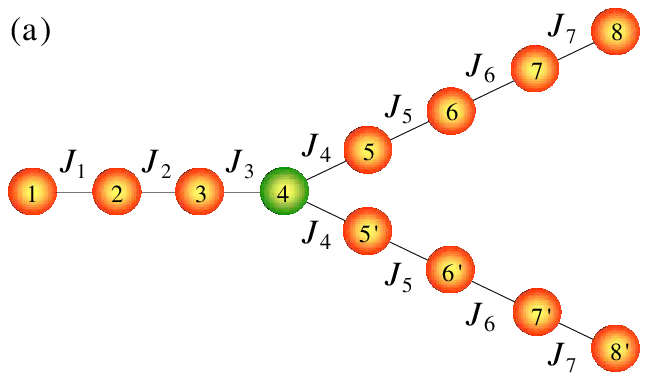}}
\resizebox{0.35\textwidth}{!}{%
\includegraphics{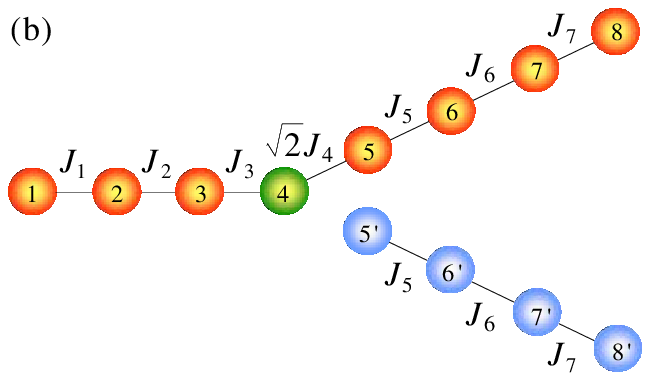}}
\caption{(Color online) (a) Sketch of the Y-shaped spin network. (b)
Under the transformation (30) the Y-shaped structure is transformed
into a linear chain consisting of the input arm and one output arm
while the other output arm is decoupled.}
\label{fig:7}       
\end{figure}

For this spin network, if we prepare initial state $|1\rangle$ in
the first node of the input arm, then after some time $t$,
entanglement will be established between the end nodes of the output
arms (for Fig.7, it corresponds to node 8 and $8'$). From the
formalism described in Section 2 one can obtain
$C(l_1,l_2,\lambda,t)=F(l,\lambda,t)$. This implies that under the
influence of intrinsic decoherence, the concurrence of the created
entanglement between the end nodes of the output arms also behaves
as a damped oscillation as the rescaled time $\lambda t$ evolves.
For infinite rescaled evolution time $\lambda t\rightarrow \infty$,
the concurrence goes to a steady state value $C^{\rm
{steady}}(l_1,l_2)=2^{2-2l}\prod_{k=2}^{l}[4-2/(k-1)]$, which can be
obtained directly from Eq. (25).

Similarly, for the multiarm structure $M(l_1,l_2,N_A)$, using the
same method, one can obtain that the concurrence measuring pairwise
entanglement between arbitrary two qubits of the end nodes of the
output arms is given by
$C(l_1,l_2,N_A,\lambda,t)=2F(l,\lambda,t)/N_A$ (when $N_A=2$, this
equality reduces to that describing the Y-shaped structure), which
observes the similar behaviors as the Y-shaped structure, i.e., it
behaves as a damped oscillation as the rescaled time $\lambda t$
evolves, and when  $\lambda t\rightarrow\infty$, it goes to a steady
value $C^{\rm
{steady}}(l_1,l_2,N_A)=2^{3-2l}\prod_{k=2}^{l}[4-2/(k-1)]/N_A$.

\section{Modified spin chains for high-fidelity state transfer}
\label{sec:5} From the above arguments one can see that the
interaction-modulated ideal spin channels for perfect state transfer
are destroyed in the presence of intrinsic decoherence environments.
Though there exists an optimal rescaled time at which one can get a
relative high transfer fidelity, however, this transfer fidelity
(including the average fidelity) decreases as the chain length $N$
increases, which puts great constrains for long distance
communication in interacting-spin systems.

Here we demonstrate that a minor modification of the exchange
interactions between the first and the last two nodes of the above
structure can fulfill the requirements of long distance and near
perfect state transfer (see Fig. 8a). To see this, we display our
numerical results for chain length $N=11$ and $N=51$ in Fig. 9(a),
from which one can see that for all decoherence rate $\gamma$, the
maximum transfer fidelity $F_{max}$ approaches unity if $J_0$ is
small enough (note that when $J_0=0$, $F\equiv0$), which indicates
that even in the presence of intrinsic decoherence environments, one
can still achieve near perfect transfer of an excitation between the
opposite ends of a XX chain by varying the strength of the exchange
interactions between the first and the last two nodes of the
modulated spin chain.
\begin{figure}
\centering
\resizebox{0.4\textwidth}{!}{%
\includegraphics{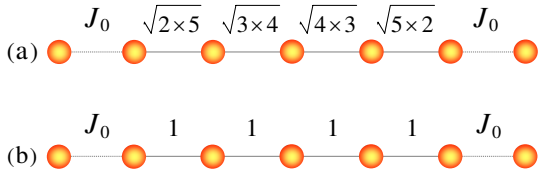}}
\caption{ (Color online) Sketches of the two modified seven-site
spin chains which may serve as spin channels for near perfect state
transfer in the presence of intrinsic decoherence environments. Here
we choose $N$ odd for it enables more efficient (i.e., high speed)
state transfer than its counterpart with even $N$.}
\label{fig:8}       
\end{figure}

Another structure which may serve as near perfect spin channel for
long distance transfer of an excitation in intrinsic decoherence
environments is the XX quantum wire with the neighboring couplings
except those between the first and the last two nodes are the same
(see Fig. 8b). This chain can serve as spin channel for an almost
perfect state transfer in the absence of decoherence environments
[11]. When the intrinsic decoherence is present, from Fig. 9(b) one
can see that a long distance transfer of an excitation whose
fidelity can be arbitrarily close to unity is also possible for very
small but nonzero $J_0$, even for large decoherence rate $\gamma$.
\begin{figure}
\centering
\resizebox{0.4\textwidth}{!}{%
\includegraphics{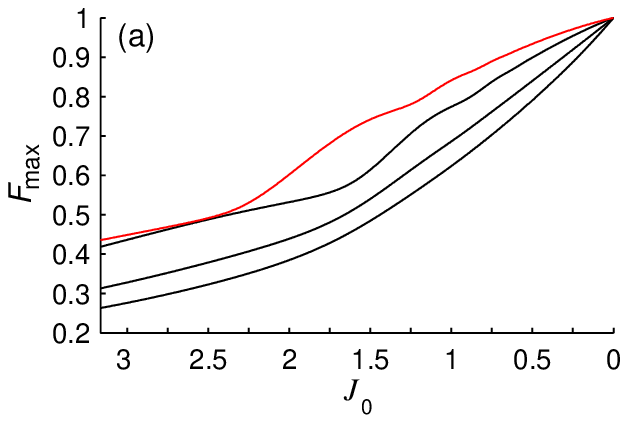}}
\resizebox{0.4\textwidth}{!}{%
\includegraphics{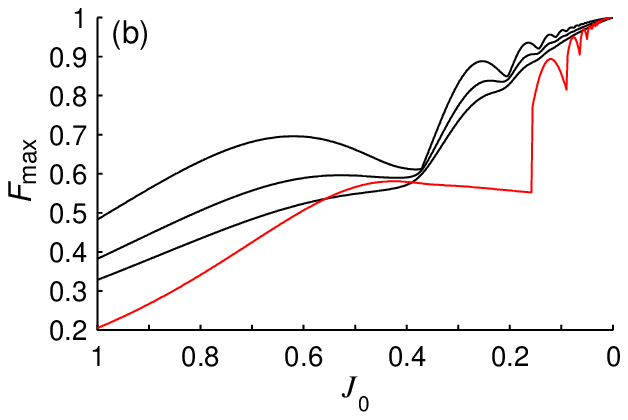}}
\caption{(Color online) Dependence of $F_{max}$ on $J_0$ for the two
modified spin channels displayed in Fig. 8. In both figures the
black lines from top to bottom correspond to chain length $N=11$ and
decoherence rate $\gamma$=0.15, 0.30 and 0.45; whereas the red lines
correspond to $N=51$ and $\gamma$=0.15. Note that when $J_0=0$,
$F\equiv0$.}
\label{fig:9}       
\end{figure}

To understand the above phenomenon, we sketch dynamics of $c_{\rm
i}(t)$ in Fig. 10, where $c_{\rm i}(t)=\sqrt{\langle
i|\rho_i(t)|i\rangle}$ denotes the amplitude of the coefficient for
the state $|i\rangle$. From these two figures one can see that for
very small but nonzero $J_0$, except the two end spins (here is node
1 and 11), the spins at the odd-number sites remain almost unexcited
during the time evolution process (i.e., $c_{\rm i}(t)|_{\rm{i\in
odd;i\neq1,N}}\simeq 0$), as if the excitation is transferred only
through the even-number nodes. In fact, a more detailed analysis
show that if $J_0\rightarrow 0^+$ and $t\leqslant t_{\rm{op}}$, the
mixedness of the system remain almost unchanged ($1-tr\rho^2\simeq
0$), i.e., the state here is very close to a pure state during the
evolution process and thus can be described approximately by
$|\psi(t)\rangle=c_1(t)|1\rangle+c_N(t)|N\rangle+\sum_{\rm{i\in
even}}c_i(t)|i\rangle$, where $c_1(t)+c_N(t)\simeq 1$ (note that
$c_1^2(t)+c_N^2(t)<1$ since $c_{\rm i}(t)|_{\rm{i\in even}}\neq 0$).

In order to better understand why the spins located at odd-number
sites (the two end spins are exceptions) remain almost unexcited
during the time evolution process for very small values of $J_0$, we
graph the effects of decreasing $J_0$ on the eigenvector population
$c_{k,1}^2=\{|\langle\hat{k}|1\rangle|^2\}_{E_k}$ for the two
modified spin chains in Fig. 11, from which one can see that with
decreasing values of $J_0$, the distribution of the eigenvectors
becomes narrower and narrower. Particularly, in the limitation
$J_0\rightarrow 0^+$, $\{|\langle\hat{k}|1\rangle|^2\}_{E_k}\neq 0$
only for the three central eigenvectors of the system, i.e.,
$c_{k,1}\neq0$ only when $k_1=(N-1)/2$, $k_2=(N+1)/2$ and
$k_3=(N+3)/2$. Moreover, for these three values of $k$, it can be
shown that when $J_0\rightarrow 0^+$, $c_{k,n}\neq0$ only when
$n=1$, $N$ and $n\in \rm{even}$ for $k=k_1$, $k_3$, and $n=1$, $N$
for $k=k_2$. For example, when $N=11$, the eigenvectors of the
Hamiltonian describing the first modified spin chain for the above
three values of $k$ can be obtained as
\begin{equation} \label{eq:32}
\left\{\begin{aligned}
  &|\hat{k}_{1,3}\rangle=\frac{1}{2}(|1\rangle+|N\rangle)\mp\sqrt{\frac{5}{42}}(|2\rangle+|10\rangle)\mp \\
  & \quad\quad\quad \sqrt{\frac{1}{12}}|6\rangle\pm \sqrt{\frac{15}{168}}(|4\rangle+|8\rangle),\\
  &|\hat{k}_{2}\rangle=\sqrt{\frac{1}{2}}(|1\rangle-|N\rangle).
        \end{aligned}\right.
\end{equation}

Similarly, for the second modified spin chain, its eigenvectors for
the above three values of $k$ with arbitrary chain length $N$
($N\in$ odd) can be obtained as
\begin{equation} \label{eq:33}
\left\{ \begin{aligned}
        &|\hat{k}_{1,3}\rangle=\frac{1}{2}[|1\rangle+(-1)^{(N+1)/2}|N\rangle]\pm \\
        & \quad\quad\quad \frac{1}{\sqrt{N-1}}\sum_{n\in\rm{even}}(-1)^{n/2}|n\rangle,\\
        &|\hat{k}_{2}\rangle=\frac{1}{\sqrt{2}}[|1\rangle+(-1)^{(N-1)/2}|N\rangle].
        \end{aligned} \right.
\end{equation}

On the other hand, for initial state $|1\rangle$ prepared in the
input node A, the density matrix at arbitrary time $t$ can be
obtained by choosing $\theta=\pi/2$, $\phi=0$ of Eqs. (10) and (11)
as
\begin{eqnarray}
\rho(t)&=&\sum_{n,m=1}^N\sum_{k,k'=1}^Nc_{k,1}c_{k',1}c_{k,n}c_{k',m}\nonumber\\&&
\exp\left[-it(E_k-E_{k'})-\frac{\gamma
t}{2}(E_k-E_{k'})^2\right]|n\rangle\langle m|.\nonumber\\
\end{eqnarray}

Combination of Eq. (34) with the above arguments, one can conclude
that the spins located at odd-number sites remain almost unexcited
during the time evolution process, i.e., the excitation is
transferred only through the even-number nodes for very small but
nonzero values of $J_0$.
\begin{figure}
\centering
\resizebox{0.4\textwidth}{!}{%
\includegraphics{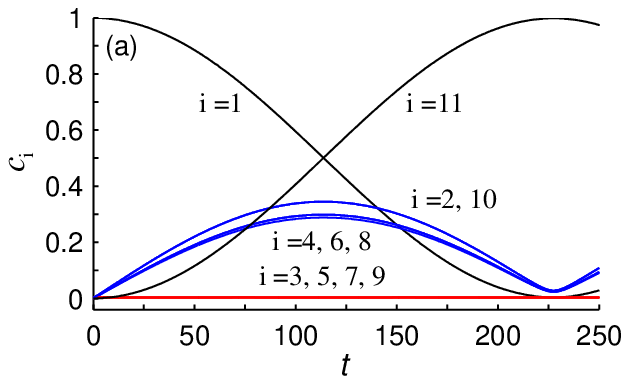}}
\resizebox{0.4\textwidth}{!}{%
\includegraphics{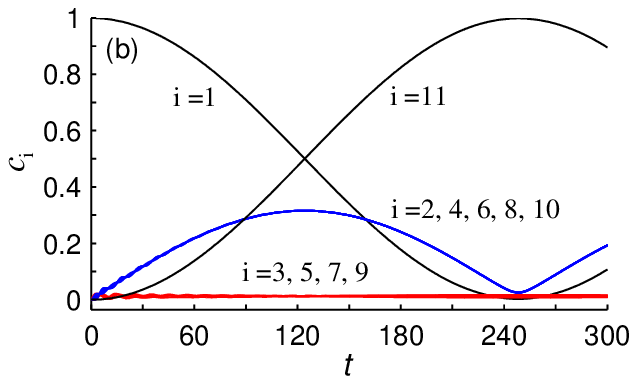}}
\caption{(Color online) Dynamics of $c_{\rm{i}}(t)$ for the two
modified spin channels displayed in Fig. 8 with $J_0=0.02$,
$\gamma=0.15$ and $N=11$. Here the curves for even-number $i$
(denoted by blue lines) are almost overlapped.}
\label{fig:10}       
\end{figure}

\begin{figure}
\centering
\resizebox{0.4\textwidth}{!}{%
\includegraphics{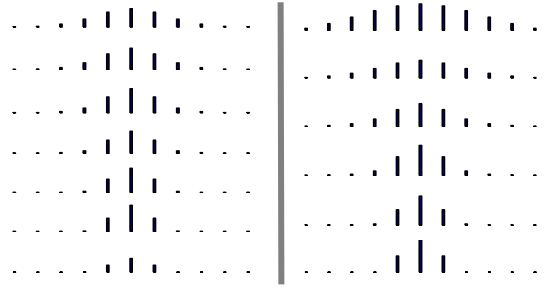}}
\caption{The eigenvector populations
$\{|\langle\hat{k}|1\rangle|^2\}_{E_k}$ (denoted by the relative
height of the vertical lines) for the system of 11 spins. The graphs
from left to right in every plot correspond to the values of $k$
increases one by one (or equivalently, the eigenvalues increase one
by one). The left seven panels correspond to the first modified spin
chain with $J_0$=3, 2.5, 2, 1.5, 1, 0.5 and $J_0\rightarrow 0^+$
(from top to bottom), while the right six panels correspond to the
second modified spin chain with $J_0$=1, 0.8, 0.6, 0.4, 0.2 and
$J_0\rightarrow 0^+$ (from top to bottom).}
\label{fig:11}       
\end{figure}

In fact, from the formalism described in Section 2, one can see that
the transfer fidelity of an excitation from one end of the chain to
another is completely determined by
\begin{eqnarray}
a_{NN}&=&\sum_{k,k'=k_1}^{k_3}c_{k,1}c_{k',1}c_{k,N}c_{k',N}\nonumber\\&&
\exp\left[-it(E_k-E_{k'})-\frac{\gamma t}{2}(E_k-E_{k'})^2\right],
\end{eqnarray}
where the coefficients of the eigenvectors for the three values of
$k$ are given by
\begin{eqnarray}
c_{k_1,1}&=&c_{k_3,1}=\frac{1}{2},\quad
c_{k_1,N}=c_{k_3,N}=\frac{(-1)^{(N+1)/2}}{2},\nonumber\\
c_{k_2,1}&=&\frac{1}{\sqrt{2}},\quad
c_{k_2,N}=\frac{(-1)^{(N-1)/2}}{\sqrt{2}}.
\end{eqnarray}

On the other hand, the eigenvalues of the two modified spin chains
correspond to the three eigenvectors
$\{|\hat{k}_1\rangle,|\hat{k}_2\rangle,$ $|\hat{k}_3\rangle\}$ can
be written as $\{-E_0,0,E_0\}$ ($E_0$ can be obtained numerically),
thus from the above two equations and the formalism described in
Section 2, one can obtain the transfer fidelity of one excitation as
\begin{eqnarray}
F&=&\frac{3}{8}+\frac{1}{8}\exp(-2\gamma
E_0^2t)\cos(2E_0t)-\nonumber\\&& \frac{1}{2}\exp\left(\frac{-\gamma
E_0^2t}{2}\right)\cos(E_0t)\quad (N\in\rm{odd}).
\end{eqnarray}

Since $E_0$ is very small in the limit of $J_0\rightarrow 0^+$, from
Eq. (37) one can see that an almost perfect transfer of an
excitation from one end of the chain to another occurs after time
$t_c\sim\pi/E_0$. Moreover, we noted that $E_0$ decreases with the
decrease of $J_0$, thus the critical time at which the transfer
fidelity $F$ gets its maximum value increases with the decrease of
$J_0$, and is independent of the decoherence rate $\gamma$. These
conclusions can be corroborated by numerical results displayed in
Fig. 10. For the system parameters adopted there (i.e., $N=11$,
$J_0=0.02$), $E_0$ can be obtained numerically as $E_{01}\simeq
0.013801$ and $E_{02}\simeq 0.012648$, thus one has $t_{c1}\sim 227$
and $t_{c1}\sim 248$. Clearly, these results agree well with those
displayed in Fig. 10.

A similar analysis shows that for the two modified spin chains with
even-number qubits, the behavior of the transfer fidelity is
determined only by the two central eigenvectors of the system in the
limit of  $J_0\rightarrow 0^+$, i.e., $c_{k,1}\neq0$ only when
$k_1=N/2$ and $k_2=N/2+1$. The eigenvectors of the Hamiltonian
describing both the two modified spin chains for the above two
values of $k$ can be obtained as
\begin{equation} \label{eq:1}
\left\{ \begin{aligned}
        &|\hat{k}_{1}\rangle=\frac{1}{\sqrt{2}}[|1\rangle+(-1)^{N/2}|N\rangle],\\
        &|\hat{k}_{2}\rangle=\frac{1}{\sqrt{2}}[|1\rangle-(-1)^{N/2}|N\rangle],
        \end{aligned} \right.
\end{equation}
with the corresponding eigenvalues $-E_0$ and $E_0$. Combination of
these with the formalism in Section 2, the transfer fidelity of one
excitation can be obtained as
\begin{eqnarray}
F=\frac{1}{2}-\frac{1}{2}\exp(-2\gamma E_0^2t)\cos(2E_0t)\quad
(N\in\rm{even}).
\end{eqnarray}

From Eq. (38) one can see that in the limit of $J_0\rightarrow 0^+$,
except the two spins located at the end nodes, all the other spins
remain almost unexcited during the time evolution process, as if the
excitation is transferred only between the two end nodes. Moreover,
from Eq. (39) one can see that an almost perfect transfer of an
excitation from one end of the chain to another occurs after time
$t_c\sim\pi/2E_0$. However, our numerical results show that the
values of $E_0$ for odd $N$ is much larger than that for even $N$
(e.g., for $J_0=0.001$, the values of $E_0$ for $N=11$ is about 2524
and 2804 times larger than that for $N=10$ and 12), thus the chain
with odd $N$ enables a more efficient (i.e., high speed) state
transfer than its counterpart with even $N$.

As a final discussion, it is worthwhile to investigate the
efficiency of the above two modified spin channels, i.e., whether
they can serve as near perfect spin channels for transfer of an
arbitrary one-qubit state by varying the strength of $J_0$. For this
purpose, we compute the average fidelity. From the above analysis
one can obtain straightforwardly that in the limit of
$J_0\rightarrow 0^+$, the average fidelity can be expressed as
\begin{eqnarray}
\bar{F}=\frac{|b_N|\cos(\alpha)}{3}+\frac{a_{NN}}{6}+\frac{1}{2},
\end{eqnarray}
where the coefficients for odd and even $N$ are given by
\begin{eqnarray}
a_{NN}&=&\frac{3}{8}+\frac{1}{8}\exp(-2\gamma E_0^2t)\cos(2E_0t)-
\nonumber\\&& \frac{1}{2}\exp\left(\frac{-\gamma
E_0^2t}{2}\right)\cos(E_0t),
\nonumber\\
b_N&=&\frac{(-1)^{(N-1)/2}}{2}\left[1-\exp\left(\frac{-\gamma E_0^2t}{2}\right)\cos(E_0t)\right],\nonumber\\
\end{eqnarray}
and
\begin{eqnarray}
a_{NN}&=&\frac{1}{2}-\frac{1}{2}\exp(-2\gamma
E_0^2t)\cos(2E_0t),\nonumber\\
b_N&=&(-1)^{N/2}i\exp\left(\frac{-\gamma
E_0^2t}{2}\right)\sin(E_0t).
\end{eqnarray}

As pointed above, $E_0$ is an infinitesimal in the limit of
$J_0\rightarrow 0^+$, thus from Eqs. (40) and (41) one can see
obviously that for odd $N$, the average fidelity approaches unity
after time $t_c\sim\pi/E_0$. For even $N$, however, due to the fact
that $\alpha=\pi/2$ is not a multiple of $2\pi$, the average
fidelity can only reach its maximum value $2/3$ (equals to the
classical average fidelity) after time $t_c\sim\pi/2E_0$. This
implies that for the above two modified spin channels with even $N$,
one cannot achieve near perfect state transfer of an arbitrary
one-qubit state simply by varying the strength of $J_0$. But if one
can apply an external magnetic field $B$ along the $z$ axis of every
spin (this does not change the eigenvectors of the system since
$[\hat{H},\sigma_{\rm{tot}}^z]=0$), the phases of the received state
at the destination node may be corrected. With this method, we
performed numerous calculations and the numerical results revealed
that the average fidelity can also approaches unity by choosing
appropriate strength of the magnetic field (e.g., for $N=10$,
$J_0=0.01$ and $B=\pm0.0003$, the two modified spin channels give
rise to $\bar{F}_{max}=0.9998$ and $\bar{F}_{max}=0.9996$,
respectively).

\section{Summary}
\label{sec:6} To summarize, we have investigated quantum state
transfer, generation and distribution of entanglement in the model
of Milburn's intrinsic decoherence environment. We focused on
diverse interaction-modulated spin networks which may serve as
perfect spin channels in the absence of decoherence. As one
expected, the state transfer fidelity as well as the amount of the
generated and distributed entanglement will be significantly lowered
by the intrinsic decoherence environment, and this detrimental
effects become severe as the decoherence rate $\gamma$ and the spin
chain length $N$ increase. For infinite evolution time $t$, we show
analytically that both the state transfer fidelity (including the
average fidelity) and the concurrence of the generated and
distributed entanglement approach steady state values, which are
independent of the decoherence rate $\gamma$. This brings great
constraints on these structures as spin channels for long distance
and high-fidelity communication. Finally, as alternative schemes to
diminish the detrimental effects, we presented two modified spin
chains which may serve as spin channels for long distance and near
perfect state transfer in the intrinsic decoherence environments.
Our results revealed that in the limit of $J_0\rightarrow 0^+$,
these two modified spin channels generate maximum fidelity 1 after
certain time $t_c\sim\pi/E_0$ for spin chains with odd-number
qubits. For spin chains with even-number qubits, however, one needs
to apply an external magnetic field in order to achieve near perfect
state transfer. \\
\\
\textbf{Acknowledgements \\} \\ This work was supported by the
National Natural Science Foundation of China under Grant No.
10547008, the Specialized Research Program of Education Bureau of
Shaanxi Province under Grant No. 08JK434, and the Youth Foundation
of Xi'an Institute of Posts and Telecommunications under Grant No.
ZL2008-11.

%

\end{document}